\documentclass[journal]{IEEEtran}

\usepackage{textcomp}
\usepackage{xcolor}
\usepackage{makecell}
\usepackage{setspace}
\usepackage{array}
\usepackage{colortbl}
\usepackage{multirow}
\usepackage{graphicx}
\usepackage{subcaption}
\usepackage{hyperref}
\usepackage{amsmath, amssymb, amsthm, cite}
\usepackage{inputenc}
\usepackage{placeins}
\usepackage{psfrag}
\usepackage{bbm}
\usepackage{bm}
\usepackage{dblfloatfix}
\usepackage{mathtools}
\usepackage{algpseudocode}
\usepackage{algorithm}
\usepackage{color}
\usepackage{tcolorbox}
\usepackage{etoolbox}
\usepackage{physics}
\usepackage{tabularx}
\usepackage{mleftright}

\allowdisplaybreaks
\newcommand{\RNum}[1]{\uppercase\expandafter{\romannumeral #1\relax}}
\usepackage{authblk}
\usepackage[english]{babel}
\usepackage{academicons}
\definecolor{orcidlogocol}{HTML}{A6CE39}

\usepackage{gensymb}

\begin{document}
\title{ Low-Resolution Radar-based Gesture Recognition}
\author{
         Netanel Blumenfeld,  
     Inna Stainvas, and
         Igal Bilik, \IEEEmembership{Senior Member, IEEE},
        \thanks{
          Authors' addresses: I. Bilik are at the School of Electrical and Computer Engineering, Ben Gurion University of the Negev, Israel. E-mail: (bilik@bgu.ac.il). This work was partially supported by the Israel Science Foundation under Grant 1895/21.
        }
    }

\maketitle
\begin{abstract}
This article addresses the challenge of recognizing hand gestures using low-resolution radar. Conventional sensors like cameras, infrared, and ultrasonic have limitations in lighting and range, while radars can provide more reliable performance. To leverage radar in gesture recognition, an innovative approach, SuperGestNet, which combines super-resolution and gesture classification networks trained jointly, is proposed. SuperGestNet consists of three sub-components: a super-resolution model that enhances radar data resolution, a pre-processing step for gesture classification, and a classification model optimized for low-power devices. Innovative pre-processing and data augmentation techniques are introduced to transform low-resolution radar data into high-resolution images using a spatially adaptive feature modulation network. This approach aims to enhance classification accuracy by generating high-quality radar images that emphasize critical features for gesture recognition. The model demonstrates improvements in both super-resolution and classification accuracy over traditional methods, making it a promising solution for reliable, low-power radar-based gesture recognition in real-world applications.

\end{abstract}

\section{Introduction}
Hand-gesture recognition (HGR) is quickly becoming a key method for human-machine interaction as it removes the need for physical contact, making access easier and more convenient. HGR involves analyzing hand movements using sensors and algorithms, allowing natural human-machine interaction without conventional input devices like a mouse or keyboard.

Applications of HGR include controlling robots \cite{malima2006fast}, home automation \cite{lu2014hand}, controlling mobile phones \cite{pu2013whole}, sign language communication for the hearing impaired \cite{pansare2012realtime}, and safer, quicker accessibility of features in automotive \cite{skaria2019interference}.

Common approaches for HGR include using camera sensors \cite{kim2008dynamic, ohnbar2014predicting, edris2017fuzzy, doermann2003progress}, infrared sensors \cite{erden2014hand}, and ultrasonic sensors \cite{sang2018micro}. Camera-based systems use computer vision algorithms to recognize gestures but require high-quality images, which can be challenging due to noisy backgrounds, lighting, and weather conditions. Infrared and ultrasonic sensors also have limitations, such as limited range and sensitivity to environmental factors. In contrast, radar sensors are not sensitive to lighting, weather conditions, or noisy backgrounds, and they are more robust in different environments. These advantages make radar a promising alternative to traditional sensors.

When using radar sensors for capturing hand gestures, the type and richness of the gesture signatures depend on the radar architecture and the employed waveform. The common types of waveforms are pulses, continuous waveform (CW), and frequency modulated continuous waveform (FMCW).

To perform radar-based HGR, several techniques have been employed, including feature extraction followed by machine learning for classification \cite{9464317, 10114956} and principal component analysis (PCA). PCA has been utilized to reduce the dimensionality of large datasets, such as Gray Level Co-occurrence Matrix (GLCM) texture features extracted from EEG spectrogram images for IQ classification \cite{mustafal2013eeg}. In another application, PCA was used to generate radiation patterns for a microwave compressive sensing system, optimizing the measurement process by requiring fewer samples compared to random-pattern-based methods \cite{liang2015reconfigurable}. Deep learning algorithms have significantly enhanced the performance of radar-based HGR by extracting Range-Doppler frames from FMCW radar and then using 2D CNN followed by LSTM \cite{wang2016interacting, 9253617}. Alternatively in \cite{scherer2020tinyradarnn} Range-Doppler frames from pulse radar are extarcted and a classifier based on a TCN network is used. Additionally, some research has focused on enhancing features from Range-Doppler maps generated from mmWave SIMO radar by using long recurrent all-convolutions neural networks (LRACN) \cite{8542778}, and by applying Clutter Extraction and CFAR algorithms on the Range-Doppler maps \cite{8662554}. Other feature extraction methods include trajectory images from UWB-radar with 2D CNN \cite{khan2020inair}. Besides different feature extraction techniques and neural network architectures, there has also been research on different learning techniques, such as meta-learning with custom loss on mmWave radar \cite{8821302} or enhancing HGR accuracy by adding micro-Doppler information \cite{8249172, 7592916, 8610109}.

The effectiveness of radar-based gesture recognition critically hinges on two types of radar resolution: range resolution and Doppler resolution. Previous works focus on high classification accuracy and robustness, selecting radar resolution according to classification task needs, thus adjusting low-cost, low-resolution radar to required classification performance. Super-resolution techniques can resolve this trade-off by replacing high-cost, high-resolution radar with low-cost, low-resolution radar enhanced by super-resolution techniques in space and frequency. There is extensive work on radar-based super-resolution techniques like classical direction of arrival (DOA) estimation, such as MUSIC and ESPRIT \cite{qi2005spatial, schmidt1986multiple, roy1989esprit}, but these methods require non-coherent sources, multiple receivers, fewer sources than receivers, and many samples for good performance. In order to make these methods more robust and applicable in real life, DOA-based deep learning has been developed \cite{deep_root_music_2023, real_time_doa_2021, resnet_doa_2022, robust_doa_2022, subspacenet_doa_2023}. However these assumptions do not hold in the case of HGR, such as the number of sources being equal to or smaller than the number of sensors. Additionally, this approach focuses on angular separation and not on distance and Doppler frequency separation.

The SR problem in computer vision has seen numerous solutions and significant advancements recently that enable training deeper and larger networks for image super-resolution \cite{8485283, 8463459}. Moreover, visual transformers (ViTs) \cite{9607618, dosovitskiy2020image, chen2021pre} outperform convolutional neural networks (CNNs) in low-level vision tasks, but ViTs are computationally expensive and inefficient for super-resolution design. Various methods have been proposed to reduce computational complexity, including efficient module design \cite{kong2022residual, hui2019lightweight, liu2020residual, zhao2020efficient, sun2022shufflemixer}, structural re-parameterization \cite{zhang2021edge}, parameter sharing \cite{ahn2018fast}, and sparse convolutions \cite{ahn2018fast}. Another direction is to improve inference time by enhancing post up-sampling techniques \cite{dong2016accelerating, shi2016real} and model quantization \cite{ignatov2022efficient}. These methods increase running time but decrease reconstruction performance, indicating room for better trade-offs between model efficiency and reconstruction performance. Some works attempt to find the sweet spot between model performance and complexity \cite{tan2021efficientnetv2, safm}. We adopt those methods to pulse-Doppler radar frames.

Super-resolution methods in computer vision aim to improve image visibility but do not necessarily add real features. For gesture classification, we need to ensure that the features from super-resolution are indeed correct. Therefore, some adjustments must be made from the domain of computer vision to our domain of recognizing gestures from radar signals. The radar signal contains features for classifying gestures, but only certain parts are relevant to this task. Thus, we focus on super-resolution that enriches and sharpens these parts while maintaining the presence of the added items. Our super-resolution aims to improve classification performance rather than general super-resolution for visibility.

The main contributions of this paper are summarized as follows:
\begin{enumerate}
    \item We developed a DL model consisting of a cascade of two networks that first reconstructs high-resolution radar measurements and then performs gesture recognition. 
    \item The networks are trained jointly using a weighed combination of super-resolution and classification losses, thus managing a trade-off between the two tasks.
    \item The ultimate high-resolution data adjusts for the gesture recognition task, enhancing its effectiveness compared with simple super-resolution.
    \item New training is explored enabling usage of shared classification network independent of the data resolution
    \item We propose and analyze a novel recursive architecture for super-resolution motivated by multi-scale theory. 
\end{enumerate}

%  
% This study seeks to demonstrate that enhanced low-resolution radar data can effectively substitute high-resolution data in gesture recognition tasks, thus offering a cost-effective and less computationally demanding alternative.

\section{Problem Definition}
This work addresses the challenges of gesture recognition and super-resolution using pulse radar with low range and Doppler resolution. 

The analytic transmitted pulse radar signal $f(t,T)$ is given by:
\begin{equation}
\begin{split}
f(t,T)=p(t-nT_P) e^{j2\pi f_c (t-nT_p)}, \forall n= 0,\ldots, M-1\;, 
\end{split}
\end{equation}
where $p(t) = A_{TX}\text{rect}\left(\frac{t-d}{d}\right)$ is the rectangular pulse with duration $d$ and amplitude $A_{TX}$, $f_c$ is the carrier frequency, $t$ is the fast time axis denoting the time within a single pulse transmission, and $T$ is the slow time axis, marking the staggered start times of successive pulses. Each pulse start is separated by a Pulse Repetition Interval (PRI) denoted by $T_{p}$, with $M$ representing the total number of transmitted pulses.

\begin{figure}[ht]
\centering
\includegraphics[width=0.45\textwidth]{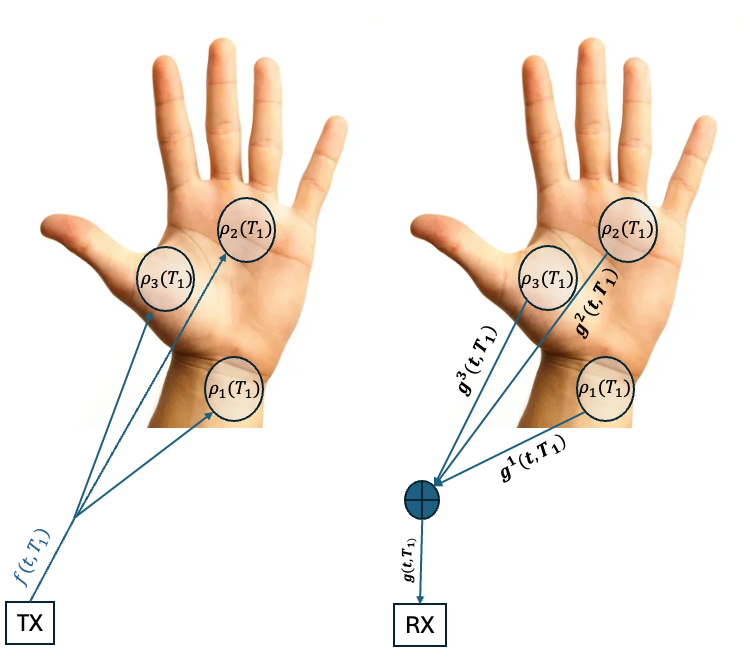}
\caption{The left figure shows the signal that send by the transmitter (TX) \(f(t, T_1)\) at time \(T_1\) and the scatter points in the hand with different cross sections \(\rho_i\). The right figure illustrates each point as a source for the reflection signal \(g^{i}(t, T_1)\). All reflected signals combine by superposition to form \(g(t, T_1)\) that receive at the receiver (RX).}
\label{fig:scatter}
\end{figure}

In the case of short-range hand gesture recognition, the hand can be considered as a distributed target consisting of a finite number of scattering centers as illustrate in Figure \ref{fig:scatter}. The scattered signal from the hand can be modeled as a superposition of multiple echo signals:
\begin{equation}
s(r, T) = \sum_{i=1}^{N_S} \rho_i(T) \delta (r - r_i(T))
\end{equation}
where $r = \frac{tc}{2}$ is the radial distance from the radar, $c$ is the speed of light, $r_i(T)$ is the radial distance from the $i$-th scatterer to the radar, $\delta (\cdot)$ is the Dirac function, $\rho_i(T)$ is the radar cross section corresponding to the $i$-th scatterer, and $N_S$ is the number of scatterers. When hand or radar move radar sensor beams hit different scattering points, so far  $r_i(T)$ and $\rho_i(T)$ will change depending on the relative pose between radar and hand.

Thus, the received signal $g(t)$ can be expressed as the convolution of the transmitted signal and the gesture scattering model:
\begin{equation}
g(t, T) = f(t, T) \otimes s(r, T)=\sum_{i=1}^{N_S} g^{i}(t,T)  
\end{equation}
where $g^{i}(t,T)$ is the received echo from the $i$-th scatterer:
\begin{equation}
g^{i}(t,T) = \frac{\rho_i(T)}{r_i^{4}(T)}p(t-T-\frac{2r_i(T)}{c})e^{j2\pi [f_c(t-T)-\frac{2r_i(T)}{c}]}
\end{equation}

Finally, the signal passes through a low-pass filter to achieve the base-band signal:
\begin{equation}
g_{lp}(t, T) = \sum_{i=1}^{N_S}\frac{\rho_i(T)}{r_i^{4}(T)}p(t-T-\frac{2r_i(T)}{c})e^{-j2\pi\frac{2r_i(T)}{c}}
\end{equation}

For each recording of a gesture, the received baseband signal is rearranged into a data cube. The data cube (denoted by $D$) contains multiple frames (the first axis), with each frame having slow and fast time axes (the second and third axes of the data cube). The data cube contains $K$ frames, with each frame denoted as $D^{k}$, where $k$ is the frame number index:
\begin{equation}
D^{k} = \begin{bmatrix}
    g_{1,1} & g_{1,2} & \cdots & g_{1,N} \\
    g_{2,1} & g_{2,2} & \cdots & g_{2,N} \\
    \vdots  & \vdots  & \ddots & \vdots  \\
    g_{M,1} & g_{M,2} & \cdots & g_{M,N}
\end{bmatrix}
\end{equation}
where $M$ is the number of received pulses (slow time) and $N$ is the number of samples of each pulse (fast time).

\section{The Proposed Approach}\label{sec:Methodology_The-Proposed-Approach}
\begin{figure}[b]
\centering
\includegraphics[width=0.45\textwidth]{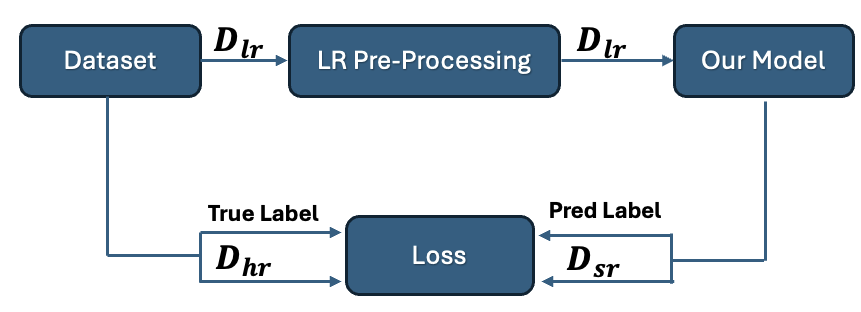}
\caption{Our proposed approach: the low-resolution data ($D_{lr}$) undergoes pre-processing and is then input into SuperGestNet, which reconstructs the data ($D_{sr}$) and the label. From $D_{sr}$, the original high-resolution data ($D_{hr}$), the true label, and the predicted label, we compute the training loss function.}
\label{fig:main_pipe}
\end{figure}

This section presents and explains the pipeline of our method, depicted in Figure \ref{fig:main_pipe}. First, the low-resolution data $D_{lr}$ undergoes pre-processing in the "LR Pre-Processing" block before entering to SuperGestNet block. SuperGestNet block illustrated in Figure \ref{fig:model_pipe} consists of three sub-blocks.

The first sub-block is the "Super-Resolution Model" block, which takes $D_{lr}$ as input and outputs $D_{sr}$, followed by the "Classification Pre-Processing" block, which processes $D_{sr}$ and outputs $\hat{D}_{sr}$ and finally, the "Classifier Model" block uses $\hat{D}_{sr}$ to perform classification. %make the classification decision.
 %The final block is the "Training Loss" block, which describes our combined SR and classification training loss.

\subsection{LR Pre-Processing}\label{subsec:Methodology_lr-pre-processing}
In the initial pre-processing stage, the focus is on preparing the low-resolution radar signal data \(D_{lr}\) for super-resolution processing. The dimensions of \(D_{lr}\) are \((F, \frac{M}{d_{s}}, \frac{N}{d_{f}})\), where \(d_s\) and \(d_f\) are the down-sampling factor applied to \(M\) and \(N\) (the fast and slow axis). The pre-processing sequence is as follows:

\begin{enumerate}
    \item \textbf{Noise Addition:} Complex Gaussian noise is added to \(D_{lr}\) to enhance the robustness of the super-resolution model by training it on a more varied dataset.
    \item \textbf{Normalization:} The data is scaled to normalized values within the $[0,1]$ range. Normalization is crucial for stabilizing the training process of neural networks by ensuring that the input values fall within a range that matches the output range of the activation functions.
    \item \textbf{Complex to Real-Imaginary Transformation:} The complex-valued data is split into its real and imaginary components. This step is performed by adding a new channel to the data cube \(R\), resulting in an output shape of \((C, F, \frac{M}{d_{s}}, \frac{N}{d_{f}})\), where \(C\) represents the two channels corresponding to the real and imaginary parts of the signal.
\end{enumerate}

% I think this is redundant for such as a short paper: This structured approach to LR data pre-processing is designed to ensure that the input to the super-resolution model is in a standardized format, facilitating more efficient and effective model training and inference.

\subsection{SuperGestNet}\label{subsec:OurModel}
\begin{figure}[b]
\centering
\includegraphics[width=0.45\textwidth]{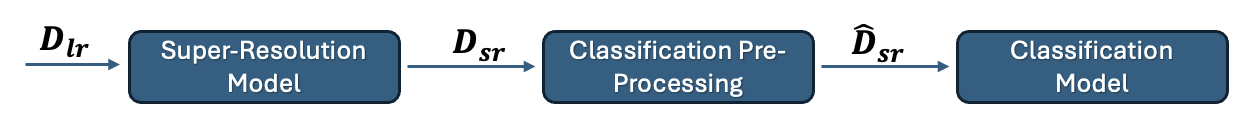}
\caption{The first step involves the super-resolution model that takes $D_{lr}$ and reconstructs $D_{sr}$. Next, $D_{sr}$ goes through a differentiable processing block to generate $\hat{D}_{sr}$, which is then input into the classifier to predict the gesture.}
\label{fig:model_pipe}

\end{figure}

Our model SuperGestNet, depicted in Figure \ref{fig:model_pipe}, consists of three integrated components: the super-resolution model, the processing stage, and the classification model. The super-resolution model first increases the resolution of \(D_{lr}\), enhancing the features responsible for classification. The high-resolution data \(D_{sr}\) is then processed to prepare it for classification, resulting in \(\hat{D}_{sr}\). Finally, the classification model categorizes the processed data into predefined classes.

The "Classification Pre-Processing" block is a differentiable block, allowing us to train our two models simultaneously as a cascade model.

\subsubsection{Super-Resolution Model}\label{subsec:SRModel}

\begin{figure}[ht]
\centering
\includegraphics[width=0.48\textwidth]{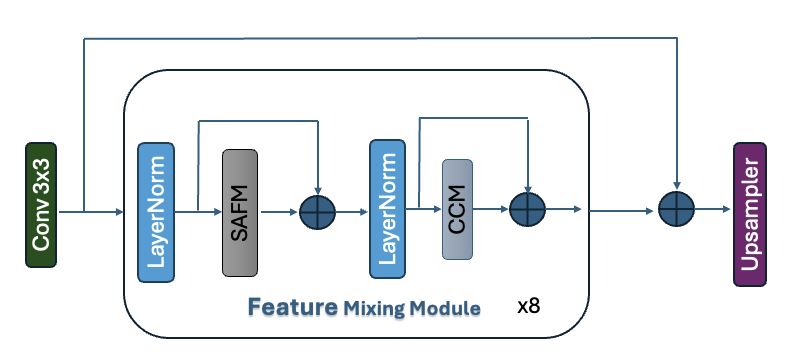}
\caption{Architecture of the Spatially-Adaptive Feature Modulation Network (SAFMN), as Super-Resolution model that used for enhancing radar image resolution through dynamic spatial modulation.}
\label{fig:safmn}
\end{figure}

Our approach utilizes the Spatially-Adaptive Feature Modulation (SAFMN) model \cite{safm}, depicted in Figure \ref{fig:safmn}. This model combines convolution processes with a spatially-adaptive feature modulation mechanism, making it highly effective for managing computational and memory constraints on low-power devices. This model is particularly beneficial for radar-based gesture recognition, where maintaining spatial details in super-resolved images is critical. The SAFMN model enhances the resolution of radar images by dynamically modulating features based on their spatial relevance.

\begin{figure}[ht]
\centering
\includegraphics[width=0.48\textwidth]{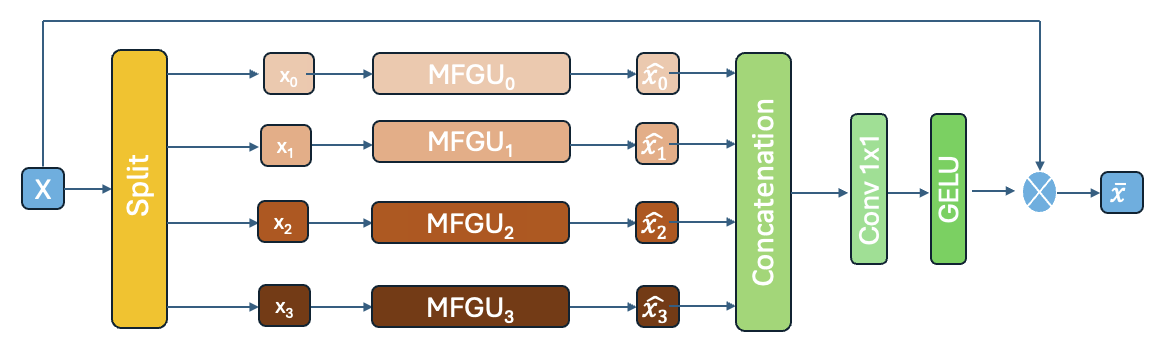}
\caption{Detailed structure of the Spatially-Adaptive Feature Modulation (SAFM) layer, showing the multi-scale processing of input features.}
\label{fig:safm}
\end{figure}

The model architecture consists of an initial 3x3 convolution layer to extract a group of base feature representation maps that can be divided into four groups for further processing. This is followed by a series of Feature Mixing Modules (FMMs) to enhance the data for high-resolution image reconstruction. Each FMM block contains a normalization layer followed by a Spatially-Adaptive Feature Modulation (SAFM) layer (detailed later). We use a residual connection by adding the input and output of the SAFM layer. The result then goes to a normalization layer and subsequently to a Channel-wise Cross Modulation (CCM) layer (detailed later), with another residual connection between the input and output of the CCM layer.

The integration of the SAFM and CCM within a unified FMM is formulated as follows:
\begin{equation}
Y = \text{SAFM}(\text{LN}(X)) + X
\end{equation}

\begin{equation}
Z = \text{CCM}(\text{LN}(Y)) + Y
\end{equation}
where LN represents layer normalization, and $X$, $Y$, and $Z$ signify the intermediate feature states.

\textbf{Spatially-Adaptive Feature Modulation (SAFM)}: The SAFM layer depicted in Figure \ref{fig:safm}, splits the input channels into four equal groups, each processed at a different scale level. This process is denoted by the Multi-Feature Grouping Unit (MFGU) layer. It includes an adaptive max pooling 2D layer to down-sample the data to the desired scale, followed by a 3x3 depth-wise convolution layer. Depth-wise convolution filters each input channel separately, making the process faster and more efficient by reducing the number of parameters and computations. After convolution, the data is interpolated back to the original dimensions. Once each group is processed at its scale, the four groups are concatenated, followed by a 1x1 convolution layer and GELU activation function. Finally, a residual connection performs element-wise multiplication between the SAFM layer's input and output to generate the feature map $\tilde{x}$.

The SAFM layer process can be formulated as follows:
\begin{equation}
\text{Split}(X) = [X_0, X_1, X_2, X_3]
\end{equation}

\begin{equation}
\hat{X}_0 = \text{DW-Conv3$\times$3}(X_0)
\end{equation}

\begin{equation}
\hat{X}_i = \uparrow_p (\text{DW-Conv3$\times$3}(\downarrow_{\frac{p}{2^i}}(X_i))), \quad i=1,2,3
\end{equation}

\begin{equation}
\hat{X} = \text{Conv1$\times$1}(\text{Concat}([\hat{X}_0, \hat{X}_1, \hat{X}_2, \hat{X}_3]))
\end{equation}

\begin{equation}
\tilde{X} = \phi(\hat{X}) \odot X
\end{equation}

The function $\text{Split}(X)$ divides $X$ into four parts, $\text{DW-Conv3$\times$3}$ refers to a $3 \times 3$ depth-wise convolution, $p$ represents the data dimensions, $\downarrow_{\frac{p}{2^i}}$ indicates an adaptive max pooling 2D layer for down-sampling with a factor of $\frac{p}{2^i}$, and $\uparrow_p$ represents the interpolation operation for up-sampling to return to the original input dimension. The MFGU block is defined by equation (11), $\phi$ is the GELU activation function, and $\odot$ denotes element-wise multiplication.

The SAFM layer is designed to enhance the representation capability of convolution by incorporating the dynamic and extensive feature interaction properties of multi-head self-attention. SAFM uses a multi-head approach, where each head processes different scale information of the input independently. The heads then combine their outputs to produce a spatially modulated attention map.

\textbf{Convolution Channel Mixer (CCM)}: The Convolution Channel Mixer (CCM) integrates local contextual information while simultaneously mixing feature channels. This module is structured around a $3 \times 3$ convolution that broadens the channel dimension to encapsulate spatial contexts, followed by a GELU activation function, and then a $1 \times 1$ convolution that compresses the channels back, ensuring efficient processing.

 \textbf{Upsampler}: The upsampler layer contains a sequence of operations to transform feature maps into an image. It first uses a convolutional layer to adjust the number of channels to match the required dimensions for upscaling. Then, it applies a PixelShuffle operation to rearrange the data into a higher resolution image based on the specified upscaling factor. The PixelShuffle operation rearranges elements in a tensor of shape $(C \times d^{2}, H, W)$ to a tensor of shape $(C, H \times d, W \times d)$, where $d$ is the upscaling factor, effectively increasing the spatial resolution of the image. In order to be able to use different down sampling scales on the fast and slow axis (\(d_{f}\) and \(d_{s}\)) we add another convolution layer with different strides to mach \(d_{f}\) and \(d_{s}\).

\subsubsection{Classification Pre-Processing}\label{subsec:classification_preprocessing}
During the pre-processing phase, the super-resolution data \(D\), initially in dimensions of \((C, K, M, N)\), is transformed back into a cube of complex values with the shape \((K, M, N)\). This restoration is crucial for the subsequent step of generating Doppler range maps. In this process, the complex-valued data \(D\) undergoes a Fast Fourier Transform (FFT) on the fast time axis, producing a two-dimensional distance-Doppler map:
\begin{equation}
\hat{D}_{sr}[f, r] = \sum_{n=0}^{M-1} D_{sr}[m, r] e^{-2\pi i f m / M}
\end{equation}

This map effectively captures the frequency shifts (Doppler effect) across different ranges, highlighting essential features of motion and position that enable precise differentiation of gestures. By transitioning from the temporal to the frequency domain, the technique accentuates unique characteristics of various hand movements, significantly improving the classifier’s ability to detect subtle differences.

\subsubsection{Classifier Model}\label{subsec:classifier_model}
The classifier architecture described in the TinyRadarNN paper \cite{scherer2020tinyradarnn} employs a temporal convolution neural network (TCN) optimized for low-power micro-controllers. This architecture is designed to handle the temporal properties of classification data effectively. Unlike Long Short-Term Memory (LSTM) networks that maintain an internal state for modeling temporal sequences, TCNs operate in a stateless manner, allowing for parallel computation of sequential outputs. This characteristic significantly reduces memory requirements for buffering feature maps, making TCNs especially suitable for deployment on embedded platforms with limited memory.

The proposed model combines a 2D Convolutional Neural Network (CNN) with a 1D TCN to address both spatial-temporal modeling and sequence modeling challenges. The 2D CNN processes input feature maps, capturing spatial and short-term temporal information within individual frames of RADAR data. Subsequently, the features computed by the 2D CNN are further processed by the TCN. The TCN utilizes an exponentially increasing dilation factor to combine features from different time steps into a single feature vector, which is then fed into a classifier consisting of fully connected layers. This approach allows the model to consider multiple consecutive output feature vectors of the 2D CNN, effectively extending the time window of analysis to capture longer-term temporal pattern.

\subsection{Training Loss Function}\label{subsec:LossFunction}
SuperGestNet's effectiveness relies on minimizing a combined loss function that includes both super-resolution quality and gesture classification accuracy. The total loss is defined as:
\begin{equation}
\text{Loss} = \gamma L_1 + L_c 
\end{equation}
Here, \(L_c\) is the classification loss, defined as the Cross-Entropy loss:
\begin{equation}
L_c = - \sum_{i=1}^{N} \sum_{c=1}^{C} y_{ic} \log(\hat{y}_{ic})
\end{equation}
where \(N\) is the number of samples, \(C\) is the number of classes, \(y_{ic}\) is a binary indicator (0 or 1) if class label \(c\) is the correct classification for sample \(i\), and \(\hat{y}_{ic}\) is the predicted probability that sample \(i\) is of class \(c\). \(L1\) represents the loss for the super-resolution task, focusing on reconstructing high-resolution radar images from low-resolution ones. 

The weighting factor \(\gamma\) is key to forcing our model to perform super-resolution in a way that enhances classification accuracy. By adjusting \(\gamma\), we can balance the influence of the super-resolution and classification tasks. 

On one hand, the classification loss forces the super-resolution process to enhance critical features for classification rather than adding details that merely improve the image's appearance. On the other hand, the \(\gamma\) factor acts as a bias factor for super-resolution, allowing us to control how much emphasis is placed on improving image quality versus classification accuracy. This careful balancing ensures that the super-resolution not only produces visually appealing images but also contributes to more accurate gesture classification.

% Each training experiment, conducted with a different \( \gamma \), will demonstrate how shifts in this balance affect the model's performance in terms of super-resolution metrics and classification accuracy. 

\section{Performance Evaluation}\label{Results}
\subsection{Evaluation Metrics}\label{subsec:EvalMetrics}
We evaluate the performance of both super-resolution and classification accuracy. The quality of super-resolved (SR) images is measured using the following metrics:

\begin{itemize}
    \item \textbf{PSNR} (Peak Signal-to-Noise Ratio): Measures the peak error between the reconstructed and original high-resolution images.
    \begin{equation}
    PSNR = 10 \cdot  \log_{10}(\frac{MAX_I^2}{\text{MSE}})  \nonumber
    \end{equation}
    
    where $MAX_I$ is the maximum possible pixel value of the data, and the MSE is defined as:
    \begin{equation}
    MSE = \frac{1}{n} \sum_{i=1}^{n} (D_{sr,i} - D_{hr,i})^2\ \nonumber
    \end{equation}

    \item \textbf{MS-SSIM}: Multi-Scale Structural Similarity Index \cite{msssimloss} assesses the visual quality of the super-resolved images at multiple scales, reflecting perceptual image quality.

    \item \textbf{L1 Loss}: Represents the mean absolute error between the predicted and true pixel values, providing a straightforward measure of average reconstruction error.
    \begin{equation}
        L_1 = \frac{1}{n} \sum_{i=1}^{n} |D_{sr,i} - D_{hr,i}| \nonumber
    \end{equation}
\end{itemize}

\subsection{Data Description}\label{sec:DataDescription}
This section provides detailed information on the datasets used in our study. It describes the source of our original high-resolution radar data and outlines the methodology employed to generate corresponding low-resolution datasets.

\subsubsection{High-Resolution Radar Data}\label{subsec:HRData}
The high-resolution radar data used in this research are from the paper
TinyRadarNN \cite{scherer2020tinyradarnn}. 
The original data were collected using low-power, short-range A1 radar sensors from Acconeer, designed to capture a variety of human gestures. This dataset includes 12 distinct classes, comprising 11 specific gestures and one "NoHand" class, recorded from 26 participants, making it highly diverse and suitable for robust gesture recognition tasks. Each recording consists of 5 frames, with the total recording time extending up to 0.625 seconds. The recording is stored as a 3D cube with dimensions (5, 32, 492), where 5 (F) is the number of frames, 32 (M) is the slow time axis, and 492 (N) is the fast time axis. This detailed temporal and spatial resolution is crucial for accurately capturing the subtle movements of human gestures.

\begin{figure*}[ht]
    \centering
    \includegraphics[width=0.98\textwidth]{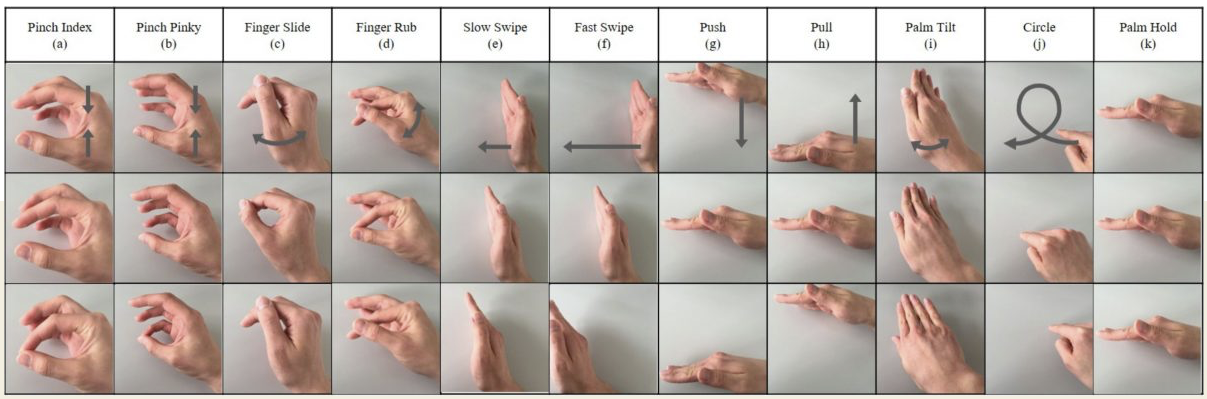}
    \caption{Overview of the 11 gestures in the dataset \cite{scherer2020tinyradarnn}}, the 12-th gesture is "NoHand"
    \label{fig:gestures}
\end{figure*}

\begin{table}[ht]
\centering
\begin{tabular}{|l|l|}
\hline
\textbf{Parameters}         & \textbf{Values}     \\ \hline
$f_c$                         & 70 GHz              \\ \hline
PRF                         & 256 Hz              \\ \hline
Sensors                     & 2                   \\ \hline
Gestures                    & 11                  \\ \hline
Recording Length            & $\leq$ 3 s          \\ \hline
Participants                & 26                  \\ \hline
Instances per Session       & 7                   \\ \hline
Sessions per Recording      & 5                   \\ \hline
Recordings                  & 20                  \\ \hline
Instances per Gesture       & 710                 \\ \hline
Instances per Participant   & 7,700               \\ \hline
Total Instances             & 7,700               \\ \hline
Sweep Ranges                & 10–30 cm            \\ \hline
Sensor Modules Used         & XR112               \\ \hline
\end{tabular}
\caption{Expanded information on the high-resolution radar data parameters.}
\label{table:hr_data_parameters}
\end{table}

\subsubsection{Generation of Low-Resolution Data}\label{subsec:LRData}
To simulate low-resolution radar data, corresponding to radar systems with reduced bandwidth (\(B\)) and lower pulse repetition frequency (\(T\)), we apply down-sampling to the original high-resolution data. The transformations applied are mathematically represented as follows:

\begin{itemize}
    \item \textbf{Reducing Bandwidth (\(B\))}: Bandwidth reduction is simulated by down-sampling the slow time axis of the radar data. If \(B_{\text{high}}\) represents the original bandwidth and \(d\) is the down-sampling factor for the slow time axis, the new bandwidth \(B_{\text{low}}\) is given by:
    \begin{equation}
        B_{\text{low}} = \frac{B_{\text{high}}}{d}
    \end{equation}
    This reduction decreases the range resolution, affecting the radar's ability to distinguish closely spaced objects.

    \item \textbf{Reducing Pulse Repetition Frequency (\(T\))}: Reducing \(T\) is achieved by down-sampling along the fast time axis. Let \(T_{\text{high}}\) be the original pulse repetition frequency, and \(d\) the down-sampling factor for the fast time axis, then the new pulse repetition frequency \(T_{\text{low}}\) is:
    \begin{equation}
        T_{\text{low}} = \frac{T_{\text{high}}}{d}
    \end{equation}
    This reduction impacts the Doppler resolution, crucial for differentiating objects based on velocity.
\end{itemize}

\begin{figure}[ht]
    \centering
    \begin{subfigure}{0.2\textwidth}
        \includegraphics[width=\textwidth]{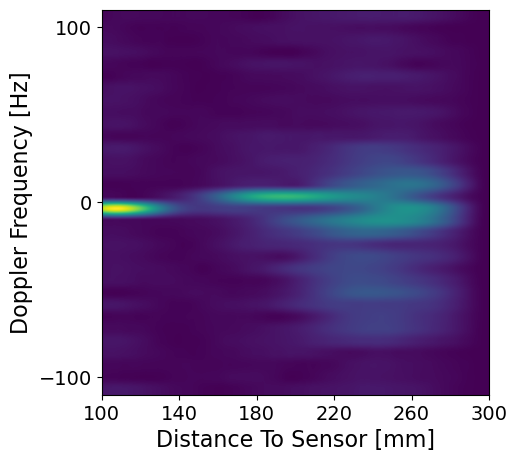}
        \caption{(a)}
        \label{fig:imgsr}
    \end{subfigure}
    \hfill  
    \begin{subfigure}{0.2\textwidth} 
        \includegraphics[width=\textwidth]{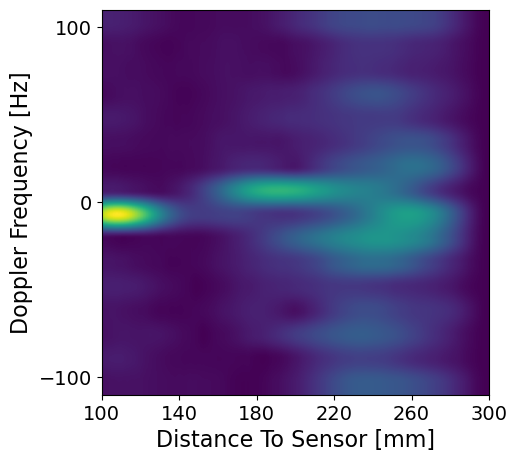}
        \caption{}
        \label{fig:imglr2}
    \end{subfigure}
    \vskip\baselineskip 
    \begin{subfigure}{0.2\textwidth}
        \includegraphics[width=\textwidth]{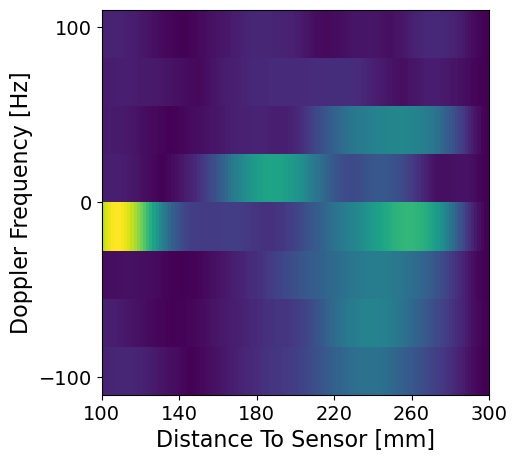}
        \caption{}
        \label{fig:imglr4}
    \end{subfigure}
    \hfill
    \begin{subfigure}{0.2\textwidth}
        \includegraphics[width=\textwidth]{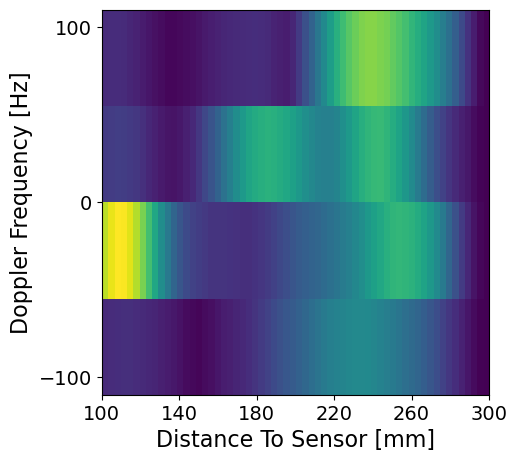}
        \caption{}
        \label{fig:imglr8}
    \end{subfigure}
\caption{\raggedright Abs Range-Doppler maps with different down-sampling factors: (a) no down-sampling, (b) factor of 2, (c) factor of 4, (d)factor of 8.}

    \label{fig:plots_ex}
\end{figure}

These down-sampling procedures are designed to mimic the operational limitations of radar with lower resolution, thus providing realistic challenges for our super-resolution algorithms. Figure \ref{fig:plots_ex} shows examples of the data converted to range-Doppler maps, down-sampled by different factors.

\subsection{Training Details}

This section provides technical details on the training process. We trained our model, SuperGestNet, denoted by $M_{\gamma}^{d}$, using different combinations of $\gamma$ and $d$, where $d = d_{f} = d_{s}$. The parameter $\gamma$, representing the weight of the super-resolution training loss, was set to 0, 0.5, 1, and 2. The parameter $d$, the down-sampling factor, was set to 2, 4, and 8. When $d = 1$, no down-sampling was performed.

We also trained our classifier with the same down-sampling factors, using cubic interpolation instead of a deep learning model, where this model name is CubicTinyRadarNN and denoted by $C^{d}$. Additionally, we trained the super-resolution model SAFMN on the data, then froze SAFMN and trained the classifier with the super-resolved data from the SAFMN model,this model name is SafmnTiny and  denoted by $FM^{d}$. Both of these models serve as benchmarks in our experiments.

\noindent Note that a different model is trained for each set of $d$ and $\gamma$ parameters.

For the super-resolution (SAFMN) model, we set the number of channels in the first $3 \times 3$ convolution layer to 36, so each scale level in the SAFM layer has 9 channels. We used 8 FFM blocks, and the up-scaling factor was set according to $d$ along both fast and slow axes.

Next, we tried to improve the classification accuracy by using 3 different SAFMN models, each one trained for a different down-sampling factor $d$, and one classifier for all three super-resolution models. This model is named MultiSuperGestNet and denoted by $SM$.

We then used SAFMN as a recursive super-resolution model, applying the model iteratively to achieve super-resolution by a factor of 2 each time. To up-sample the image by a factor of 4, we applied SAFMN twice before classification. This model is named RecSuperGestNet and denoted by $RM$.

For RecSuperGestNet and MultiSuperGestNet we used $\gamma = 2$ and the same parameters for SAFMN like before.

% This section provides technical details on the training process. We trained our model SuperGestNet, denoted by $M_{\gamma}^{d}$, using different combinations of $\gamma$ and $d$. The parameter $\gamma$, which represents the weight of the super-resolution training loss, was set to 0, 0.5, 1, and 2. The parameter $d$, the down-sampling factor, was set to 2, 4, and 8. When $d = 1$, no down-sampling was performed. We also trained our classifier with the same down-sampling factors without using the super-resolution network, considering this as a benchmark, where $\gamma$ is not applicable and was denote by *.
% Note, that a different model is trained for each set of $d$ and $\gamma$ parameters.

% TODO

% For the super-resolution model, we set the number of channels in the first 3x3 convolution layer to 36, so each scale level in the SAFM layer has 9 channels, and we used 8 FFM blocks.

\subsection{Down-sample Effect on Classification Accuracy}
\begin{figure}[ht]
\centering
\includegraphics[width=0.48\textwidth]{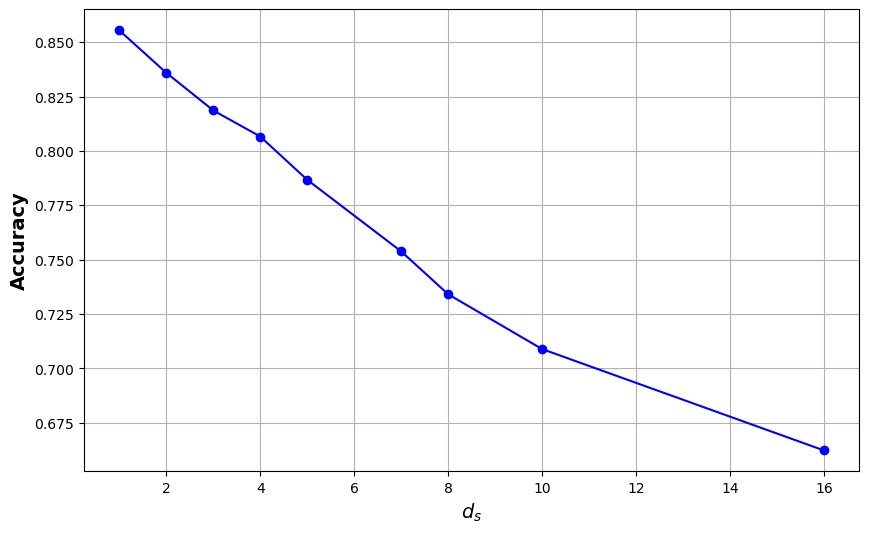}
\caption{slow time vs acc.}
\label{fig:slow_time}
\end{figure}

\begin{figure}[ht]
\centering 
\includegraphics[width=0.48\textwidth]{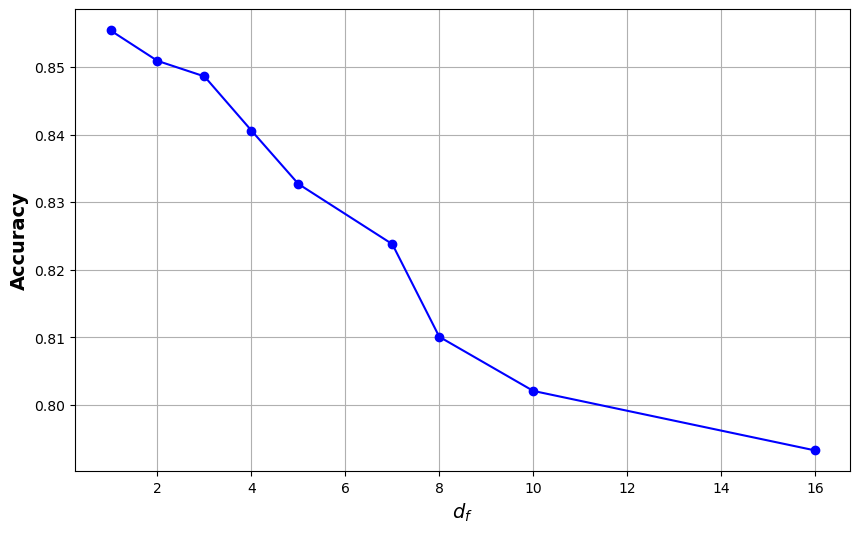}
\caption{fast time vs acc.}
\label{fig:fast_time}
\end{figure}

To understand how lower resolution impacts classification accuracy in both distance (fast time axis) and Doppler (slow time axis), we trained $C_{d}$ on lower resolution data.
First, we examined the effect of Doppler resolution on classifier performance by setting $d_f=1$ and varying $d_s$. The decrease in accuracy is shown in Figure \ref{fig:slow_time}. Next, we assessed the impact of distance resolution by setting $d_s=1$ and varying $d_s$. The decrease in accuracy is shown in Figure \ref{fig:fast_time}. As can be seen, the Doppler resolution has a greater impact on classification accuracy compared to distance resolution, this indicates that features from the Doppler axis are more critical for classification. Additionally, the results emphasize the sensitivity of the classification to the resolution of the data and the need to develop a super resolution method that can highlight the features relevant to the classification.


\subsection{Performance of the Super-Resolution}

The purpose of super-resolution (SR) is to extract important features to enhance classification accuracy. Each model uses a different approach for SR. CubitTinyRadarNN ($C^{d}$) employs traditional 'cubic' interpolation, SafmnTiny ($FM_{d}$) uses a pre-trained SR model (SAFMN), and SuperGestNet ($M_{d}^{\gamma}$) is trained to perform SR and classification together as a cascade model.

In the first two models, the SR task is performed separately from the classification task, so classification does not impact the SR results. In contrast, $M_{d}^{\gamma}$ is trained with a training loss that balances between SR and classification tasks. The classification loss can be considered a constraint that guides SR model to enhance relevant features for classification accuracy.
Table \ref{table:performance_metrics} present the results of the super-resolution metrics of $M_{\gamma}^{d}$, $FM^{d}$ , $C^{d}$ under different settings of $\gamma$ and $d$. For any value of $d$ the model $FM^{d}$ outperforms $M_{\gamma}^{d}$ for all the $\gamma$ values on the super-resolution metrics.
This is expected because  $FM^{d}$ uses an SR model trained solely on the SR task, while $M_{\gamma}^{d}$ is trained for SR with a classification constraint. The effect of $\gamma$ on the SR result is clear, a higher $\gamma$ means higher priority for the SR task, which leads to better performance in the SR task.

\begin{figure}
    \centering
    \begin{subfigure}{0.2\textwidth}
        \includegraphics[width=\textwidth]{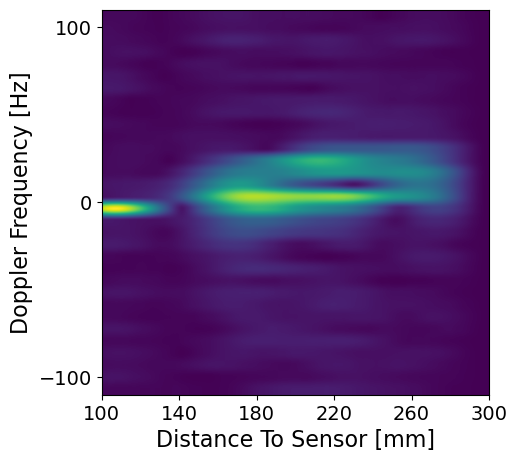}
        \caption{}
        \label{fig:imgsr11}
    \end{subfigure}
    \hfill  
    \begin{subfigure}{0.2\textwidth} 
        \includegraphics[width=\textwidth]{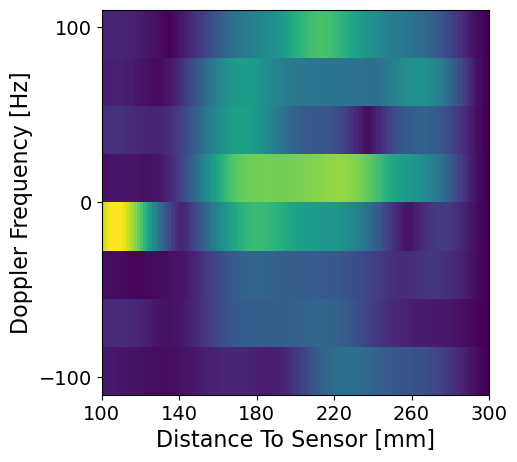}
        \caption{}
        \label{fig:imglr21}
    \end{subfigure}
    \vskip\baselineskip 
    \begin{subfigure}{0.2\textwidth}
        \includegraphics[width=\textwidth]{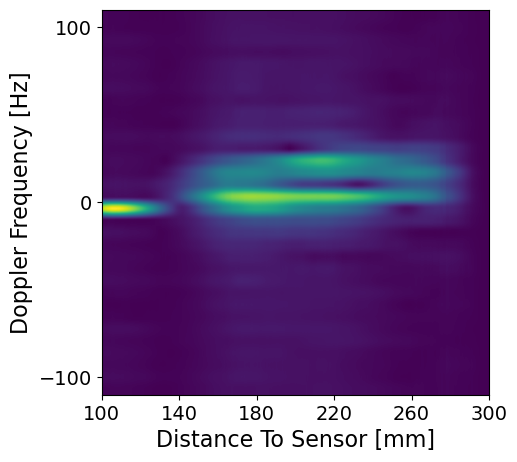}
        \caption{}
        \label{fig:imglr41}
    \end{subfigure}
    \hfill
    \begin{subfigure}{0.2\textwidth}
        \includegraphics[width=\textwidth]{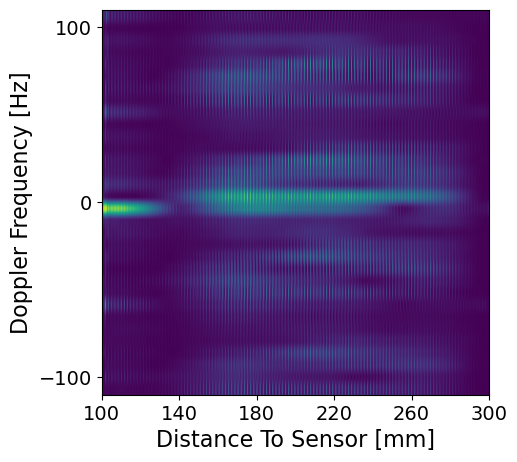}
        \caption{}
        \label{fig:imglr81}
    \end{subfigure}
    \caption{Abs Range-Doppler maps (a) original high resolution data, (b) doopler map that generated from down sample data by factor of 4, (c) super-resolution doopler map that generated with $FM^{4}$, (d) super-resolution doopler map that generated with $M^{4}_{2}$.}
    \label{fig:plots_ex_sr}
\end{figure}

An example of SR is shown in Figure \ref{fig:plots_ex_sr}. First, we apply SR to the data, then perform FFT on the fast axis and take the absolute value to obtain the range-Doppler map. The result of $FM^{4}$ is closer to the original data compared to  $M^{4}_{2}$.

\subsection{Effect of Gamma on Classification Accuracy}

Table \ref{table:performance_metrics} presents the classification accuracy of $C^{d}$,  $FM^{d}$, and  $M_{\gamma}^{d}$ under various down-sampling factors $d$ and $\gamma$ values. $C^{d}$ and $FM^{d}$ serve as benchmark models to assess the impact of Super-Resolution (SR) on improving classification accuracy across down-sampling factors $d=d_f=d_s=1, 2, 4, 8$ and to highlight the difference between jointly and separately training the super-resolution model and the classifier. The primary focus is on classification accuracy, which is crucial for evaluating the effectiveness of each model.

For $d = 1$, $C^{1}$ achieves an accuracy of 83.95\%. This serves as the baseline for comparison when no down-sampling is applied.

When $d = 2$, the accuracy results show variations among different models. $C^{2}$ accuracy slightly decreases to 81.85\%.  $FM^{2}$, which incorporates a pre-trained SR model, achieves an accuracy of 83.16\%, indicating its capability to recover important features lost due to down-sampling.  $M_{\gamma}^{2}$ models exhibit a range of accuracies depending on the $\gamma$ value. $M_{2}^{2}$ achieves 84.00\%, while the best performance is observed with  $M_{1}^{2}$, reaching an accuracy of 85.20\%. This suggests that balancing SR and classification tasks (with $\gamma = 1$) can improve accuracy. However, when $\gamma$ is reduced further, $M_{0.5}^{2}$ still performs well at 84.80\%, while $M_{0}^{2}$ shows a significant drop to 82.28\%, highlighting the importance of SR in enhancing classification accuracy.

As $d = 4$ increases to 4, the accuracy of all models decreases, reflecting the greater challenge posed by higher down-sampling. $C^{4}$ accuracy drops to 77.93\%. $FM^{4}$ maintains relatively high accuracy at 79.16\%.  $M_{\gamma}^{4}$ models show a similar trend to $d = 2$, with  $M_{0.5}^{4}$ achieving the highest accuracy at 81.46\%, followed by $M_{2}^{4}$ and  $M_{1}^{4}$ with 81.34\% and 80.91\%, respectively.  $M_{0}^{4}$ again performs poorly with an accuracy of 78.20\%.

For $d = 8$ , the accuracy further declines due to the increased difficulty of the task.  $C^{8}$ reaches 74.10\%. $M_{\gamma}^{8}$ models show varying performance with the highest accuracy from $M_{0.5}^{8}$ at 76.78\%. $M_{2}^{8}$ and $M_{1}^{8}$ achieve 76.62\% and 76.04\%, respectively. The performance of $M_{0}^{8}$ drops significantly to 73.35\%, indicating the necessity of SR for maintaining classification accuracy at higher down-sampling levels.

In summary, for each down-sampling factor, SuperGestNet ($M_{\gamma}^{d}$) outperforms SafmnTiny ( $FM^{d}$) in classification accuracy, although  $FM^{d}$ achieves better performance on the SR task. This emphasizes the advantage of joint training of the tasks, with classification used as regularization for SR. Additionally, we see a dependency of the $\gamma$ value on the down-sampling factor. As the down-sampling factor increases, $\gamma$ decreases. This is because the corruption of the data is greater when the down-sampling factor increases, and SR cannot completely recover all the missing data. In such cases, a smaller $\gamma$ will give less weight to the SR task.

\hspace{-1.5 cm}
\begin{table}[ht]
\centering
\caption{Performance Metrics for CubitTinyRadarNN, SafmnTiny and  SuperGestNet with different down-sampling factors d and different $\gamma$ values }
\label{table:performance_metrics}
\small
\begin{tabular}{|p{1cm}|p{1.2cm}|p{1cm}|p{1.5cm}|p{1cm}|}
\hline
\textbf{Model} & \textbf{Accuracy} & \textbf{L1} & \textbf{MS-SSIM} & \textbf{PSNR} \\ \hline
\multicolumn{5}{|c|}{\cellcolor[gray]{0.9}\textbf{d = 1}} \\ \hline
$C^{1}$ & 83.95 & - & - & - \\ \hline
\multicolumn{5}{|c|}{\cellcolor[gray]{0.9}\textbf{d = 2}} \\ \hline
$C^{2}$ & 81.85 & 0.215 & 0.756 & 14.93 \\ \hline
$FM^{2}$ & 83.16 & 0.033 & 0.936 & 21.7 \\ \hline
$M_{2}^{2}$ & 84.00 & 0.065 & 0.871 & 19.18 \\ \hline
\textbf{$M_{1}^{2}$} & \textbf{85.20} & \textbf{0.074} & \textbf{0.852} & \textbf{18.4} \\ \hline
$M_{0.5}^{2}$ & 84.80 & 0.093 & 0.817 & 16.98 \\ \hline
$M_{0}^{2}$ & 82.28 & 1.167 & 0.342 & -3.45 \\ \hline
\multicolumn{5}{|c|}{\cellcolor[gray]{0.9}\textbf{d = 4}} \\ \hline
$C^{4}$ & 77.93 & 0.274 & 0.711 & 13.87 \\ \hline
$FM^{4}$ & 79.16 & 0.061 & 0.876 & 19.93 \\ \hline
$M_{2}^{4}$ & 81.34 & 0.085 & 0.857 & 17.46 \\ \hline
$M_{1}^{4}$ & 80.91 & 0.093 & 0.847 & 16.93 \\ \hline
\textbf{$M_{1}^{4}$} & \textbf{81.46} & \textbf{0.108} & \textbf{0.833} & \textbf{15.77} \\ \hline
$M_{0}^{4}$ & 78.20 & 1.355 & 0.294 & -5.54 \\ \hline
\multicolumn{5}{|c|}{\cellcolor[gray]{0.9}\textbf{d = 8}} \\ \hline
$C^{8}$ & 74.10 & - & - & - \\ \hline
$M_{2}^{8}$ & 76.62 & 0.117 & 0.826 & 15.48 \\ \hline
$M_{1}^{8}$ & 76.04 & 0.134 & 0.818 & 14.36 \\ \hline
\textbf{$M_{0.5}^{8}$} & \textbf{76.78} & \textbf{0.167} & \textbf{0.794} & \textbf{12.43} \\ \hline
$M_{0}^{8}$ & 73.35 & 0.45 & 0.578 & -10.96 \\ \hline
\end{tabular}
\end{table}

\subsection{Training Networks Separately}
The purpose of this section is to highlight the difference between SafmnTiny and SuperGestNet. Although both models share the same architecture, the main difference lies in their training processes and the applied loss functions.
To train SafmnTiny, we first trained the super-resolution (SR) model, SAFMN, using the L1 loss. After training, we froze the model and proceeded to train the classifier separately with Cross Entropy loss, utilizing the super-resolved data generated by SAFMN. These two training steps were performed independently, so neither training phase influenced the other.
Training a super-resolution model typically aims to create high-resolution images that appear visually appealing and realistic. However, standard super-resolution training does not guarantee that the generated high-resolution data accurately reflects the original content. This can lead to artifacts or added features that may not have existed in the original image but are imperceptible to the human eye.
In contrast, SuperGestNet employs a cascade training approach where both models are trained jointly. The combined loss function, as defined in equation (15), integrates the L1 loss for super-resolution and the Cross Entropy loss for classification. By balancing these components with the parameter $\gamma$, this approach ensures that the super-resolution model enhances only the features necessary for accurate classification while constraining it from adding non-existent features to the data. This joint training strategy aligns the super-resolution task more closely with the classification objective, leading to more relevant and reliable outputs.

In Figure \ref{fig:plots_ex_sr}we can observe the original Doppler map (a), the Doppler map generated from down-sampled data by a factor of 4 on both axes (b), the Doppler map produced by SafmnTiny (c), and the Doppler map generated by SuperGestNet (d). The results from SafmnTiny appear more realistic and closer to the original data compared to those from SuperGestNet, which is attributed to the distinct training processes and the absence of classification loss in SafmnTiny. Notably, most of the data in the Doppler map is zero, indicating that enhancing the entire image is unnecessary. Focusing on the relevant parts helps improve classification performance and avoids introducing modifications that do not contribute to the task.

Table \ref{table:performance_metrics} shows that the super-resolution metrics for SafmnTiny are superior to those for SuperGestNet, indicating that SafmnTiny performs better on the entire dataset. However, the classification accuracy of SuperGestNet surpasses SafmnTiny's across all values of $d$ and $\gamma$. This suggests that while SafmnTiny excels in generating visually accurate high-resolution images, SuperGestNet’s joint training approach enhances the features relevant for classification, resulting in better performance in classification tasks.

The comparison between SafmnTiny and SuperGestNet underscores the trade-offs between separate and joint training strategies. While SafmnTiny produces superior overall image quality, SuperGestNet's cascade approach effectively balances image quality and classification accuracy. This demonstrates that, for applications where classification performance is critical, integrating classification loss into the training process can yield better results by tailoring super-resolution enhancements to task-specific requirements.

\subsection{MultiSuperGestNet Model}

The purpose of the super-resolution model in SuperGestNet is to enhance the relevant features for classification. The input dimensions for the classifier remain constant, and the performance metrics of the super-resolution are similar for down-sampling factors of 2 and 4, as shown in Table \ref{table:performance_MultiSuperGestNet}. This similarity indicates that the difference between the super-resolved data for down-sampling factors of 2 and 4 is not significant.

Given these observations, we propose MultiSuperGestNet, which incorporates three SR models corresponding to down-sampling factors of 2, 3, and 4, along with a single classifier. To train MultiSuperGestNet, we utilized our dataset and, for each batch, sampled the data with different down-sampling factors. For each batch, if the data was sampled with a down-sampling factor of 2, the SR model for factor 2 was chosen; if the data was sampled with a down-sampling factor of 3, the SR model for factor 3 was chosen; similarly, for a down-sampling factor of 4, the SR model for factor 4 was chosen and we set $\gamma = 2$ for all super-resolution models. This training method effectively increased our dataset size by a factor of three because each data sample produced three training samples—one for each down-sampling factor. 

The results in Table \ref{table:performance_MultiSuperGestNet} indicate an improvement in classification accuracy for each down-sampling factor. This is expected, as the single classifier is trained on an augmented dataset that is effectively three times larger. While there is a slight decrease in the SR metrics, it is not substantial. Furthermore, MultiSuperGestNet reduces memory usage by consolidating the weights of 3 classifiers into a single model, allowing it to handle three different resolutions efficiently.

Table \ref{table:performance_MultiSuperGestNet} presents the results for MultiSuperGestNet. While there is a slight decrease in the super-resolution metrics, an improvement in classification accuracy is observed.

\begin{table}[h]
    \centering
    \caption{MultiSuperGestNet Performance}
    \label{table:performance_MultiSuperGestNet}
    \begin{tabular}{|p{1.2cm}|c|c|c|c|}
        \hline
        Down-sampling Factor & Accuracy & L1 & MS-SSIM & PSNR (dB) \\
        \hline
        2 & 86.41 & 0.085 & 0.896 & 16.08 \\ \hline
        3 & 85.68 & 0.1 & 0.86 & 14.92 \\ \hline
        4 & 84.86 & 0.125 & 0.8 & 13.69 \\ 
        \hline
    \end{tabular}
\end{table}

\subsection{RecSuperGestNet Model}

To optimize memory usage, we explored training a single SR model capable of recursively increasing the resolution by a factor of 2. This approach allows us to apply the SR block once for a down-sampling factor of 2, twice for a down-sampling factor of 4, and three times for a down-sampling factor of 8. We refer to this model as RecSuperGestNet. By using a single SR block for multiple down-sampling factors, RecSuperGestNet reduces memory requirements, replacing the need for separate SuperGestNet models for each factor and consolidating the classifier into one unified structure.

We trained RecSuperGestNet for down-sampling factors of 2, 4, and 8. The results, shown in Table \ref{table:performance_RecSuperGestNet}, indicate that for a down-sampling factor of 2, the accuracy (85.61\%) is comparable to that of $M^{2}_{1}$. For a down-sampling factor of 4, the classification accuracy drops by approximately 1\% to 80.51\%, demonstrating a slight trade-off in performance. For a down-sampling factor of 8, the classification accuracy decreases further to 74.77\%, and the SR metrics (L1, MS-SSIM, and PSNR) show a noticeable decline. This reflects the increased challenge of maintaining super-resolution quality as the recursion depth grows and highlights the diminishing returns in accuracy when dealing with higher down-sampling factors.

\begin{table}[h]
    \centering
    \caption{Performance Metrics for RecSuperGestNet}
    \label{table:performance_RecSuperGestNet}
    \begin{tabular}{|p{0.4cm}|c|c|c|c|}
        \hline
        d & Accuracy (\%) & L1 Loss & MS-SSIM & PSNR (dB) \\
        \hline
        2 & 85.61 & 0.087 & 0.807 & 16.2 \\ \hline
        4 & 80.51 & 0.125 & 0.65 & 13.58 \\ \hline
        8 & 74.77 & 0.181 & 0.59 & 11.27 \\ 
        \hline
    \end{tabular}
\end{table}

These results suggest that RecSuperGestNet effectively balances memory efficiency and performance, especially for lower down-sampling factors. The model maintains strong classification accuracy and reasonable SR metrics for $d=2$, showcasing its capability to perform comparably to standard methods. While there is a decline in performance for $d=4$, the model still offers a good balance between memory savings and accuracy. However, for $d=8$, the notable drop in both classification accuracy and SR quality indicates that recursive application becomes less effective, emphasizing the limitations of this approach at higher down-sampling factors.

\subsection{Impact of Dataset Size on Model Performance}

\begin{figure}
    \centering
    \begin{subfigure}{0.2\textwidth}
        \includegraphics[width=\textwidth]{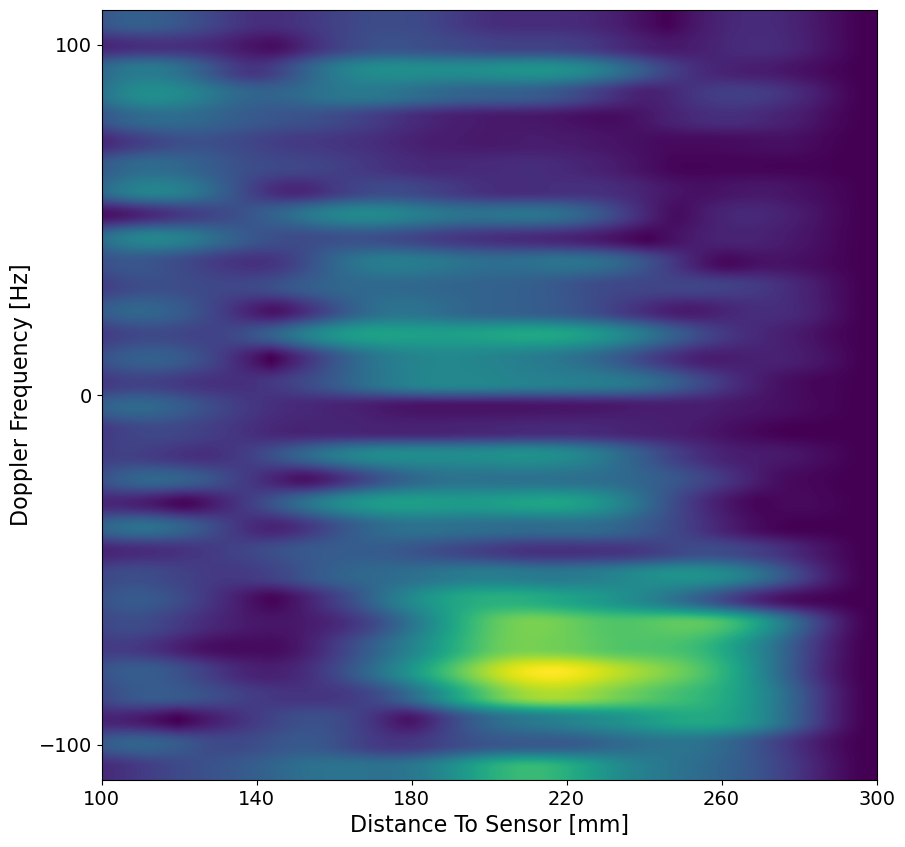}
        \caption{}
        % \label{fig:imgsr11}
    \end{subfigure}
    \hfill  
    \begin{subfigure}{0.2\textwidth} 
        \includegraphics[width=\textwidth]{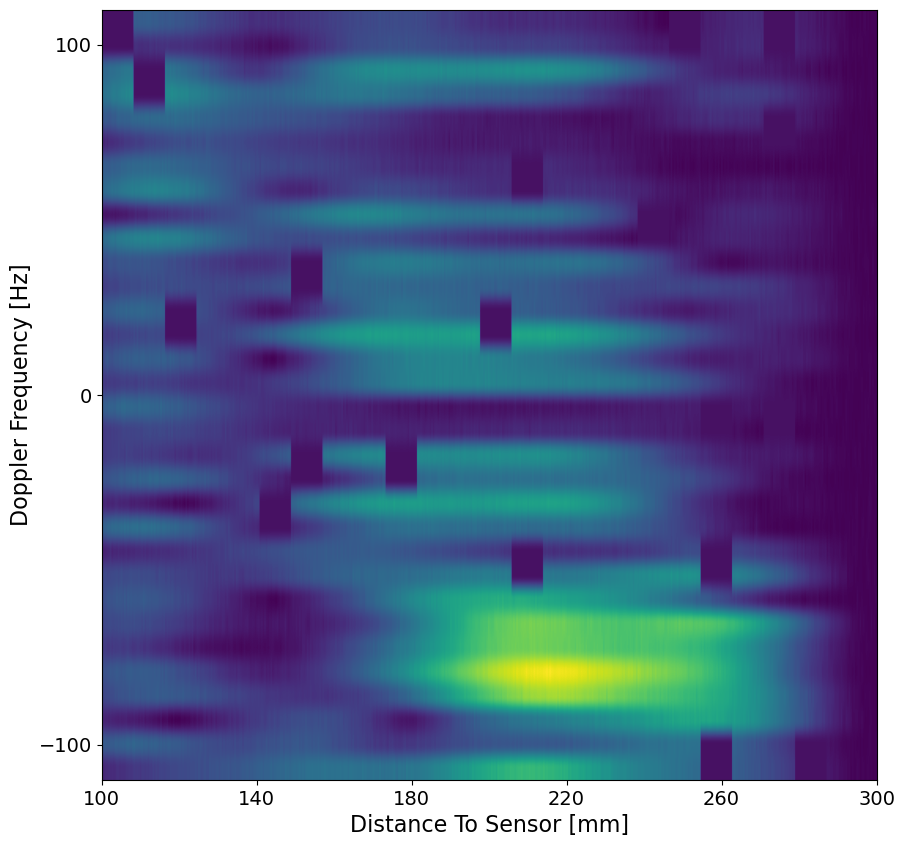}
        \caption{}
        % \label{fig:imglr21}
    \end{subfigure}
        \vskip\baselineskip 
    \begin{subfigure}{0.2\textwidth}
        \includegraphics[width=\textwidth]{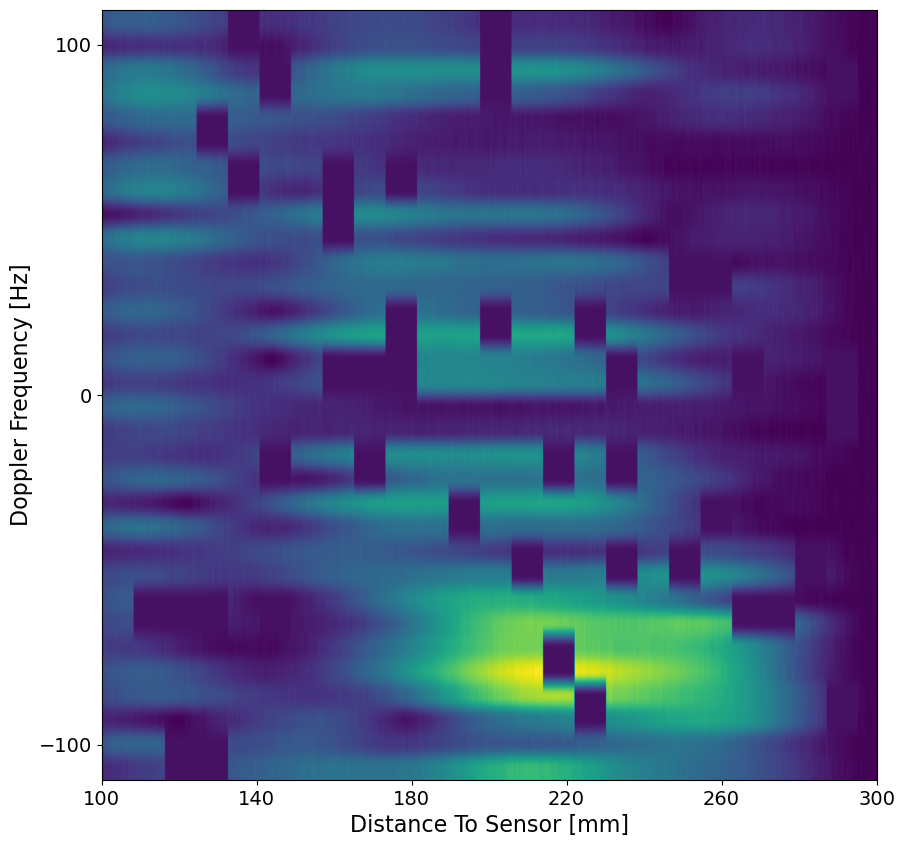}
        \caption{}
        \label{fig:imglr41}
    \end{subfigure}
    \hfill
    \begin{subfigure}{0.2\textwidth}
        \includegraphics[width=\textwidth]{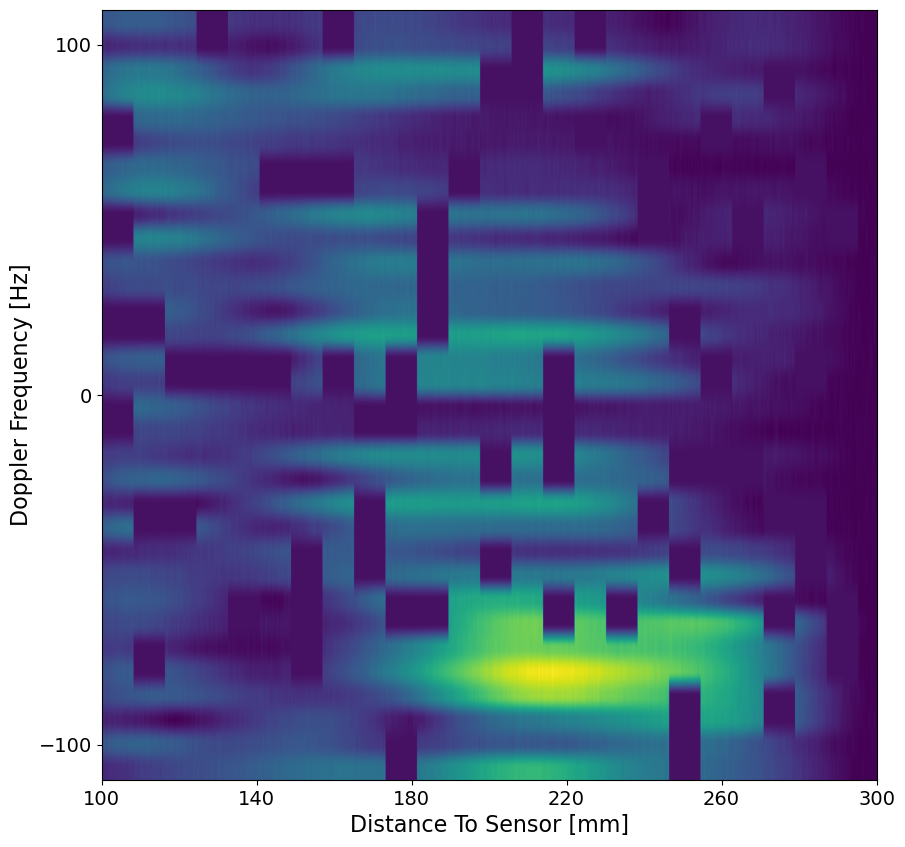}
        \caption{}
        \label{fig:imglr81}
    \end{subfigure}
    \caption{Abs of the signal in the time domain, (a) is the original signal, (b) the signal with $p=5\%$, (c) the signal with $p=10\%$, (d) the signal with $p=15\%$}
        \label{fig:augmentation}
\end{figure}

To investigate whether the improvement in classification accuracy for MultiSuperGestNet and RecSuperGestNet was solely due to the increased amount of training data, we conducted an additional experiment. We applied data augmentation techniques to expand the training dataset and trained the original SuperGestNet model using only data with a down-sampling factor of 4. The augmentations included adding Gaussian noise and a patch-based masking method based on \cite{augmentations}. Specifically, we divided each image into regular, non-overlapping patches and sampled a subset of patches while masking (i.e., removing) the remaining ones. The sampling strategy was straightforward: random sampling without replacement, following a uniform distribution. This approach, referred to as "random sampling," effectively created a task with a high masking ratio that minimized redundancy and challenged the model by preventing solutions based solely on extrapolation from neighboring patches. The use of a uniform distribution also mitigated potential center bias, ensuring that masked patches were evenly distributed across the image.
Figure \ref{fig:augmentation} present an  example of the augmentation over different percentage of block random sample (p).

Despite the augmented dataset size being comparable to that of MultiSuperGestNet and RecSuperGestNet, table \ref{table:performance_metrics_augmentation} shows that the accuracy of the augmented SuperGestNet model was lower than that of the other two models. This result indicates that the improved accuracy observed in MultiSuperGestNet and RecSuperGestNet is not solely due to the larger dataset size. Instead, it suggests that the training strategies inherent to MultiSuperGestNet and the recursive nature of RecSuperGestNet contribute significantly to their enhanced classification performance. These approaches effectively leverage the structure of the models and the training process to better align the super-resolution task with the classification objective, leading to more effective feature enhancement. While the augmented SuperGestNet model showed some improvement in accuracy, MultiSuperGestNet and RecSuperGestNet still outperformed it, demonstrating their superior classification accuracy.

\begin{table}
    \centering
    \caption{Performance Metrics for $M^{4}_{2}$ with different random sampling percentage (p).}
    \label{table:performance_metrics_augmentation}
    \begin{tabular}{|p{0.5cm}|c|c|c|c|}
        \hline
        p & Accuracy & L1 & MS-SSIM & PSNR (dB) \\
        \hline
        0 & 81.34 & 0.085 & 0.857 & 17.46 \\ \hline
        5 & 81.41 & 0.078 & 0.844 & 17.54 \\ \hline
        10 & 81.48 & 0.72 & 0.836 & 18.12 \\ \hline
        15 & 80.81 & 0.93 & 0.88 & 16.3 \\ 
        \hline
    \end{tabular}
\end{table}

% \subsection{params search}
% table \ref{table:params_search_dim}
% \begin{table}[h]
%     \centering
%     \caption{Performance Metrics for SuperTinyRadar with $d_{f} = 4, d_{s} = 2 $}
%     \label{table:params_search_dim}
%     \begin{tabular}{|p{1.2cm}|c|c|c|c|}
%         \hline
%         Dims & Accuracy & L1 & MS-SSIM & PSNR (dB) \\
%         \hline
%         4 & 80.11 & 0.113 & 0.756 & 5.076 \\ \hline
%         8 & 80.87 & 0.11 & 0.77 & 13.34\\ \hline
%         16 & 81.71 & 0.097 & 0.837 & 14.91 \\ \hline
%         32 & 81.3 & 0.089 & 0.847 & 15.34 \\ \hline
%         64 & 80.82 & 0.083 & 0.862 & 16.2 \\ \hline
%         96 & 80.79 & 0.076 & 0.871 & 16.58 \\ 
%         \hline
%     \end{tabular}
% \end{table}

\subsection{Model Performance and Memory Usage Analysis}

The comparison of model performances provides insights into trade-offs between inference speed, classification accuracy, and memory usage, especially from the perspective of practical deployment. TinyRadarNN, functioning as a standalone classifier, has the fastest inference time (0.008 seconds) and the lowest parameter count (45,948), making it an efficient choice for real-time applications that prioritize speed over comprehensive feature enhancement. However, in scenarios where more detailed classification is needed, models incorporating super-resolution (SR) components offer significant advantages.

SuperGestNet have inference time of 0.179 seconds and a parameter count of 272,162, reflecting its added complexity aimed at enhancing input features for better classification. MultiSuperGestNet extends this approach by incorporating three SR models for down-sampling factors of 2, 3, and 4 with a shared classifier. Despite its higher parameter count (735,586), if only one resolution is needed, MultiSuperGestNet can be configured with a single SR model, making its parameter count comparable to that of SuperGestNet.

From a memory usage perspective, MultiSuperGestNet consolidates three classifiers into one shared model structure, optimizing memory and inference efficiency across different resolutions. This design is beneficial for manufacturers and clients who may need flexibility in hardware, as the gesture recognition block operates on a CPU/GPU processor independent of the radar itself. This means that, in scenarios involving multiple radars, a single gesture recognition model can be shared instead of requiring separate classifier instances for each radar. For cases where only one resolution is processed, MultiSuperGestNet can be reduced to a configuration with a single SR model, making its memory usage and parameters similar to SuperGestNet, thereby maintaining flexibility without increasing the computational load unnecessarily.

RecSuperGestNet, designed to optimize memory by using a recursive SR approach, offers a significant reduction in memory requirements by employing a single SR block for various resolutions. It performs well at lower down-sampling factors but exhibits a decline in classification accuracy and SR quality at higher factors due to the increased challenge of recursive application. This makes it suitable for memory-constrained environments where multi-resolution capabilities are necessary but with an understanding of the trade-offs in performance.
.

table \ref{table:model_comparison}

\begin{table}[h!]
\centering
\begin{tabular}{|p{2.7cm}|p{2.5cm}|p{2.1cm}|}
\hline
\textbf{Model Name} & \textbf{Inference Time (sec)} & \textbf{Model Prams} \\ \hline
TinyRadarNN & 0.008245 & 45,948 \\ \hline
SuperGestNet & 0.179208 & 272,162 \\ \hline
MultiSuperGestNet & 0.134241 & 735,586 \\ \hline
RecSuperGestNet & 0.144764 & 272,144 \\ \hline
\end{tabular}
\caption{Comparison of Model Inference Time and Total Parameters}
\label{table:model_comparison}
\end{table}

\section{Conclusions}
This work demonstrates that improved hand gesture recognition can be effectively achieved using low-resolution radar data by applying deep learning techniques for super-resolution and classification. The proposed SuperGestNet model successfully enhances radar image resolution, allowing it to extract detailed spatial features critical for distinguishing gestures. By integrating a super-resolution network and a specialized gesture classification module, SuperGestNet overcomes the inherent limitations of low-resolution radar, achieving performance that rivals higher-resolution systems. This model also demonstrates practicality for real-world applications, with an architecture optimized for low-power devices, ensuring efficiency in deployment contexts. The findings highlight the potential of radar for reliable gesture recognition under challenging environmental conditions, positioning SuperGestNet as a viable alternative to traditional sensing technologies like cameras and infrared sensors, particularly in low-visibility or variable lighting scenarios. Future work will focus on refining the model for even broader gesture sets and adapting the approach for other low-resolution radar-based applications, further broadening the scope of radar in human-computer interaction fields.

\end{document}